\documentclass[twocolumn,showpacs,amsmath,amssymb,superscriptaddress,aps]{revtex4-2}
\usepackage[dvips]{graphics}
\usepackage{epsfig}
\usepackage{amsmath,bm}
\usepackage{color}

\newcommand{\car}{\mbox{$^{16}$C}}
\newcommand{\carn}{\mbox{$^{14}{\rm C}+n$}}
\newcommand{\carnn}{\mbox{$^{14}{\rm C}+n+n$}}
\newcommand{\card}{\mbox{$^{16}{\rm C}+d$}}
\newcommand{\carp}{\mbox{$^{16}{\rm C}+p$}}
\newcommand{\emax}{\mbox{$E_{\rm max}$}}

\newcommand{\jmax}{\mbox{$j_{\rm max}$}}
\newcommand{\kmax}{\mbox{$K_{\rm max}$}}
\newcommand{\elab}{\mbox{$E_{\rm lab}$}}
\newcommand{\ecm}{\mbox{$E_{\rm c.m.}$}}

\begin{document}

\title{Breakup effects in the $\carp$ and $\card$ reactions}

\author{Shubhchintak}
\email{shubhchintak@ulb.be}
\author{P. Descouvemont}
\email{pierre.descouvemont@ulb.be}
\affiliation{Physique Nucl\'eaire Th\'eorique et Physique Math\'ematique, Code Postal 229,
Universit\'e Libre de Bruxelles (ULB), B 1050 Brussels, Belgium}
\date{\today}

\begin{abstract}
We analyze the $\carp$ and $\card$ reactions within the four- and five-body Continuum Discretized
Coupled Channel (CDCC) method. The $\car$ nucleus is described by a $\carnn$ configuration in
hyperspherical coordinates. This description reproduces fairly well several $\car$ low-lying states.
First we analyze the $2^+\rightarrow 0^+$ $E2$ transition amplitude, which confirms that an effective
charge must be introduced to reproduce the experimental value.
Then, proton and deuteron elastic and inelastic scattering are investigated by including $\car$ pseudostates, which
simulate the $\carnn$ continuum. In $\card$, the deuteron breakup is taken into account with $p+n$
two-body pseudostates. A fair agreement with experiment is obtained without any fitting parameter. Breakup
effects are in general small, but improve the agreement with experiment.  
\end{abstract}
\maketitle

\section{Introduction}
\label{sec1}
The structure of light exotic nuclei has been intensively studied in recent years \cite{TSK13}. Neutron-rich
nuclei, located near the neutron dripline, are expected to present unusual properties. In
particular, the $\car$ nucleus has attracted much attention, owing to a small
E2 transition probability between the ground state and the $2^+$ first excited state (see Ref.\ \cite{Fo16}
for a review of recent works).

Since exotic nuclei are usually characterized by a short lifetime, their experimental study requires radioactive
beams, and the theoretical interpretation of the data is based on reaction models. A well-established
framework is the Continuum Discretized Coupled Channel (CDCC) method, which is well suited to exotic
nuclei since it includes the continuum of the projectile. Exotic nuclei being weakly bound, the
continuum plays an important role, even for elastic scattering.

A recent experiment \cite{JLY20} aims at measuring $\carp$ and $\card$ elastic scattering, as well as inelastic
scattering. These data complement previous experiments involving heavy targets \cite{EDK04}. The $\carp$
and $\card$ systems
can be studied theoretically within the CDCC method, where $\car$ is described by a three-body
$\carnn$ structure. The extension of the CDCC method to three-body projectiles is recent \cite{MHO04}, and
this method has been even extended to two-body targets such as the deuteron \cite{De20}. The calculations
are very time-consuming, but can be performed with modern computing facilities, and optimized codes.

The text is organized as follows. In Sec.\ \ref{sec2}, we present the $\car$ description in a $\carnn$
three-body model. We discuss more specifically the $2^+ \rightarrow 0^+$ transition probability which has
been measured, and calculated previously \cite{HS06}. Section \ref{sec3} is devoted to a brief presentation
of the CDCC theory, and of the resolution of the (large) coupled-channel system. In Secs.\ \ref{sec4} and
\ref{sec5}, we discuss the $\carp$ and $\card$ elastic scattering, respectively. Inelastic scattering is
analyzed in Sec.\ \ref{sec6}. Concluding remarks and outlook are presented in Sec.\ \ref{sec7}.

\section{Three-body model of $\car$}
\label{sec2}
\subsection{Hyperspherical method}
We use the hyperspherical coordinates to describe the three-body structure of $\car$ which we consider as made up of  a $^{14}$C core and of two valence neutrons, i.e.\ as a $\carnn$ system. Here we give an outline of the hyperspherical method and the reader is referred to Refs.\ \cite{ZDF93,Li95,DDB03} for more detail. 

In our approach, we neglect the internal structure of $^{14}$C and interactions among the three two-body systems are considered. Considering $A_1$ and $Z_1e$ as the mass number and charge of the core, we adopt the Jacobi coordinates ($\pmb{x}, \pmb{y}$) as
\begin{align}
{\pmb x} = \frac{1}{\sqrt{2}}({\pmb r}_3-{\pmb r}_2) \,\,\,\, {\pmb y} = \sqrt{\frac{2A_1}{A_1+2}}\Big({\pmb r}_1-\frac{{\pmb r}_2 + {\pmb r}_3}{2}\Big), 
\label{e1}
\end{align}
which represent one of the three possible sets of Jacobi coordinates (see for example Refs.\ \cite{ZDF93,Li95}). This choice also ensures the symmetry of the wave functions with respect to the two-neutron exchange. In
Eq.\ (\ref{e1}), ${\pmb r}_i$ are the coordinates of the core and of the neutrons, respectively. 

The hyperradius $\rho$ and the hyperangle $\alpha$ are then defined as
\begin{align}
\rho = \sqrt{x^2 + y^2} ,\hspace*{1cm} \alpha = {\rm arctan}\Big(\frac{y}{x}\Big),   
\label{e2}
\end{align}
where $\alpha$ varies from 0 to $\pi/2$.
In these coordinates, the Hamiltonian of $\car$ can be written as
\begin{align}
H_0 = T_\rho + \sum_{i<j} V_{ij}(x_k), 
\label{e3}
\end{align}
where $V_{ij}$ represent two-body potentials ($\carn$ and $n+n$) and the kinetic energy $T_{\rho}$ is given by
\begin{align}
T_\rho = -\frac{\hbar^2}{2m_N}\Big(\frac{\partial^2}{\partial \rho^2}+ \frac{5}{\rho} \frac{\partial}{\partial \rho}-\frac{K^2(\Omega_{5\rho})}{\rho^2}\Big), 
\label{e4}
\end{align} 
with $\Omega_{5\rho}= (\Omega_x, \Omega_y, \alpha)$. In this definition, $m_N$ is the nucleon mass and $K^2$ is the five-dimension angular momentum which has eigenvalues $K(K+4)$ and eigenfunctions
\begin{align}
{\mathcal Y}^{\ell_x \ell_y}_{KL M_L}(\Omega_{5\rho}) = \phi^{\ell_x \ell_y}_K(\alpha)\big[Y_{\ell_x}(\Omega_x) \otimes Y_{\ell_y}(\Omega_y) \big]^{LM_L}, 
\label{e5}
\end{align}
where $\ell_x$ and $\ell_y$ are the orbital momenta associated with ${\pmb x}$ and ${\pmb y}$. 
The hyperradial function $\phi^{\ell_x \ell_y}_K(\alpha)$ is given by
\begin{align}
\phi^{\ell_x \ell_y}_K(\alpha)  = N^{\ell_x \ell_y}_K ({\rm cos}\, \alpha)^{\ell_x} ({\rm sin}\, \alpha)^{\ell_y} P_n^{\ell_y +\frac{1}{2}, \ell_x +\frac{1}{2}}({\rm cos}\,2\alpha), 
\label{e6}
\end{align}
where $N^{\ell_x \ell_y}_K$ is a normalization factor (see for example Ref.\ \cite{DDB03}) and $P_n^{\ell_y +\frac{1}{2}, \ell_x +\frac{1}{2}}(x)$ is a Jacobi polynomial with the positive integer $n$ given by 
\begin{align}
	n=(K-\ell_x-\ell_y)/2.
\end{align}

Equation (\ref{e5}) can be extended by introducing the spinor $\chi^{S M_S}$ ($S = 0$ or 1) to take into account the spin of the external neutrons. A spin mixing is possible when the two-body interactions contain
a spin-orbit term.
We define
\begin{align}
{\mathcal Y}^{jm}_{\gamma K}(\Omega_{5\rho}) = \big[ {\mathcal Y}^{\ell_x \ell_y}_{K L}(\Omega_{5\rho}) \otimes \chi^S \big]^{jm}, 
\label{e7}
\end{align}
where index $\gamma$ is defined as $\gamma=(\ell_x, \ell_y, L, S)$ and $j$ is the total angular momentum.

The three-body wave functions corresponding to Hamiltonian (\ref{e3}) can be written as
\begin{align}
\Psi^{jm\pi}=\rho^{-5/2} \sum_{K=0}^{\infty}\sum_{\gamma} \chi_{\gamma K}^{j\pi}(\rho)\,
{\mathcal Y}^{jm}_{\gamma K}(\Omega_{5\rho}), 
\label{e8}
\end{align}
where $\pi$ stands for the parity. In practice, the summation over $K$ is truncated at some value $\kmax$. In Eq.\ (\ref{e8}), the hyperradial functions $\chi_{\gamma K}^{j\pi}(\rho)$ are obtained by solving the set of coupled differential equations
\begin{align}
&\Big(-\frac{\hbar^2}{2m_N}\Big[\frac{d^2}{d\rho^2}-\frac{(K+3/2)(K+5/2)}{\rho^2}\Big]-E^{j\pi}\Big)\chi_{\gamma K}^{j\pi}(\rho) \nonumber\\
&\hspace*{1cm}+ \sum_{\gamma' K'}V_{\gamma' K',\gamma K}^{j\pi}(\rho) \chi_{\gamma' K'}^{j\pi}(\rho) = 0, \label{e9}
\end{align}
where the coupling potentials $V_{\gamma' K',\gamma K}^{j\pi}(\rho)$ represent the matrix elements of the two-body potentials in Eq.\ (\ref{e3}) between hyperspherical functions (\ref{e7}) (see Refs.\ \cite{ZDF93,DDB03}).
The three-body energies $E^{j\pi}$ are defined from the $\carnn$ threshold.

We solve Eq.\ (\ref{e9}) by using the Lagrange-mesh method \cite{Ba15,PDB12,DDC15}, which permits fast and accurate numerical computations. The square-integrable solutions of Eq.\ (\ref{e9}) are obtained by expanding the hyperradial functions over $N$ Lagrange basis functions \cite{Ba15} $u_i(\rho)$ as
\begin{align}
\chi_{\gamma K}^{j\pi}(\rho)= \sum_{i = 1}^N  c_{\gamma K i}^{j\pi}u_i(\rho), 
\label{e10}
\end{align}
where $c_{\gamma K i}^{j\pi}$ are the expansion coefficients. For more detail, we refer to Refs.\ \cite{PDB12,DDB03}.

\subsection{Energy levels of $\car$}
\label{eng16C}
As it is clear from the previous discussion, the $n+n$ and $\carn$ two-body potentials are important inputs in our calculations. For the former, we adopt the central part of the Minnesota potential with the exchange parameter $u=1$ \cite{TLT77}. The $\carn$ potential is taken from Ref.\ \cite{HS06} (set B), which also reproduces the low-lying energy spectrum of $^{15}$C. This potential contains  forbidden states in the $s_{1/2}$, $p_{1/2}$ and $p_{3/2}$ partial waves. We remove these forbidden states by using a supersymmetric (SS) transformation \cite{BAY87}. We use $N = 20$ Gauss-Laguerre basis functions and $K_{\rm max}=20$. Numerical tests indicate that
these values are sufficient to achieve an excellent convergence.

\begin{figure}[ht]
	\centering
	\includegraphics[width=6.5cm]{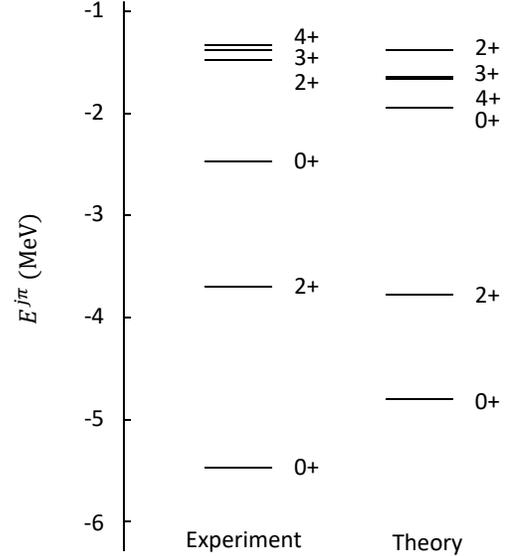}
	\caption{Energy spectrum of $\car$ [$E^{j\pi}$ values in Eq.\ (\ref{e9})]. The experimental data are taken from Ref.\ \cite{TWC93}. }
	\label{spectrum}
\end{figure}

In Fig.\ \ref{spectrum}, we compare the calculated energies of the first low-lying states of $^{16}$C with their experimental value. Apart from slight differences for the $0^+$ states, one can see that the calculated energies are quite close to the experimental values. In particular, the $2^+_1$ state is
well reproduced by the three-body model. As the ground state is deeply bound, its precise energy is not expected to be important in scattering calculations. We therefore do not include a phenomenological three-body force to
compensate for the slight difference between theory and experiment.
The three-body bound and pseudostate wave functions of $\car$ obtained in this way are then used as an
input of the CDCC calculations.

\subsection{E2 transition}
The $B(E2,2^+\rightarrow 0^+)$ transition probability has been measured in several experiments 
\cite{IOA04,EDK04,OIS08,EAF08,WFM08,PPC12,JLY20}, with results ranging from 0.63 $e^2.{\rm fm}^4$ to
4.34 $e^2.{\rm fm}^4$. A small value is consistent with the shell-model picture, where 4 protons are
in a closed $0p_{3/2}$ subshell, and 2 neutrons in the $0d_{5/2}$ subshell. Large values, however,
suggest core-polarization effects. Calculations in the shell-model \cite{YX12,KM20} and in the 
$\carnn$ three-body model \cite{HS06} require significant effective charges to reproduce the experimental
$B(E2,2^+\rightarrow 0^+)$ value. A review of recent experiments and calculations can be found in Ref.\ 
\cite{Fo16}.

\begin{figure}[ht]
	\centering
		\includegraphics[width=8.6cm, clip]{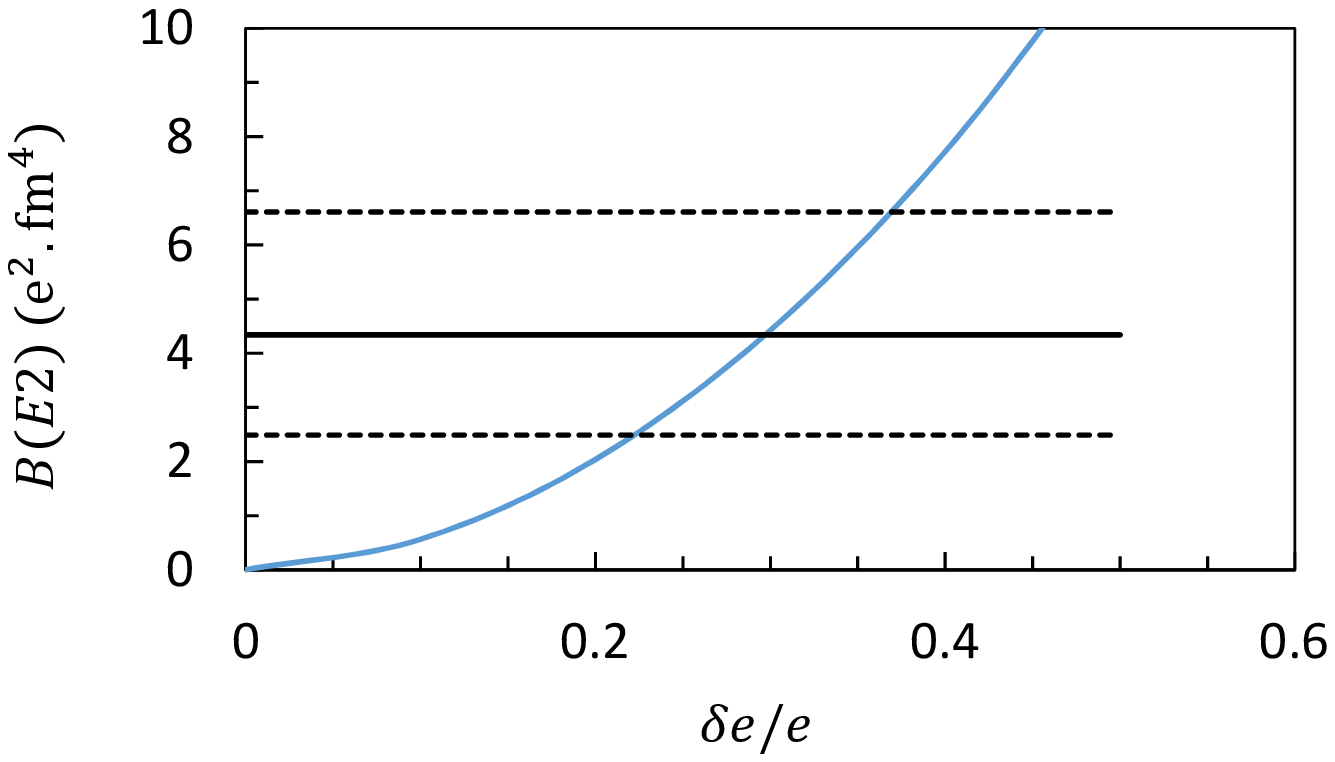}
	\caption{$E2$ transition probability for the $B(E2,2^+\rightarrow 0^+)$ transition in $\car$.
		The horizontal lines represent the latest experimental value \cite{JLY20} (with the
		dashed lines as lower and upper limits). }
	\label{be2}
\end{figure}

The three-body wave function (\ref{e8}) can be used to determine the $E2$ transition probability.
The $B(E2)$ between an initial
state $J_i\pi_i$ and a final state $J_f\pi_f$ is defined as
\begin{align}
B(E2,J_i\pi_i\rightarrow J_f\pi_f)=\frac{2J_f+1}{2J_i+1}\bigl\vert (e+\delta e)M_p + \delta e M_n \bigr\vert^2.
\label{pd1}
\end{align}
where $\delta e$ is the effective charge.
For the system considered here (a core surrounded by two neutrons), the proton and neutron matrix elements are given by
\begin{align}
	& M_p=Z_1\bigl(\frac{2}{A}\bigr)^2 \langle \Psi^{J_f\pi_f}\Vert {\cal M}_2(\pmb{y}) \Vert \Psi^{J_i\pi_i}
		\rangle, \nonumber \\
		&M_n=\langle \Psi^{J_f\pi_f}\Vert \frac{1}{2} {\cal M}_2(\pmb{x})+
		\frac{4N_1+2A_1^2}{A^2} {\cal M}_2(\pmb{y}) \Vert \Psi^{J_i\pi_i}\rangle,
	\label{pd2}
\end{align}
with the multipole operators
\begin{align}
	& {\cal M}_{2\mu}(\pmb{x})=2x^2Y_2^{\mu}(\Omega_x), \nonumber \\
	&{\cal M}_{2\mu}(\pmb{y})=\frac{A}{2A_1}y^2Y_2^{\mu}(\Omega_y).
	\label{pd3}
\end{align}
The model provides $M_p=0.173$ fm$^2$ and $M_n=14.93$ fm$^2$. The $B(E2)$ is displayed in Fig.\
\ref{be2} as a function of the effective charge $\delta e/e$. This curve is similar to the results
obtained by Horiuchi and Suzuki \cite{HS06}. The latest experimental value $4.34^{+2.27}_{-1.85}$
\cite{JLY20} is represented as horizontal lines. Without effective charge, the theoretical $B(E2)$ 
value is close to zero. Reproducing the experimental value requires $\delta e/e \approx 0.3\pm 0.1$.

\section{Outline of the CDCC theory}
\label{sec3}
The CDCC method is well adapted to investigate reactions involving weakly bound nuclei \cite{Ra74,KYI86,AIK87,YOM12}. It was originally developed to study $d$+nucleus scattering \cite{Ra74} and has been found successful in explaining the data of many reactions involving the deuteron. Actually, due to the low breakup threshold of the exotic nuclei, it becomes important to take into account their breakup effects. In the CDCC method, these effects are simulated by approximating the continuum by pseudostates (PS) which correspond to positive eigenvalues of the Schr{\"o}dinger equation associated with the projectile or/and with the
target.

Earlier applications of this method were mainly dealing with typical two-body projectiles such as $d$, $^7$Li, $^{11}$Be on structureless targets \cite{KYI86,AIK87}. However, it is now possible to study the scattering of three-body projectiles such as $^{6}$He, $^{9}$Be, $^{11}$Li \cite{MHO04,DDC15,De20} and also systems involving a two-body projectile and a two-body target such as $^{11}$Be + d \cite{De17,De18}. Recently, in Ref.\ \cite{De20}, the CDCC method has been used to study the $^{11}$Li + $d$ scattering within a 3+2 body model. In the
present paper, we follow the same formalism to study the $\carp$ and $\card$ scattering considering them as 3+1 and 3+2 body systems, respectively.

\begin{figure}[ht]
\centering
\includegraphics[width=0.55\textwidth,trim=3.0cm 2.0cm 11.0cm 1.0cm,clip]{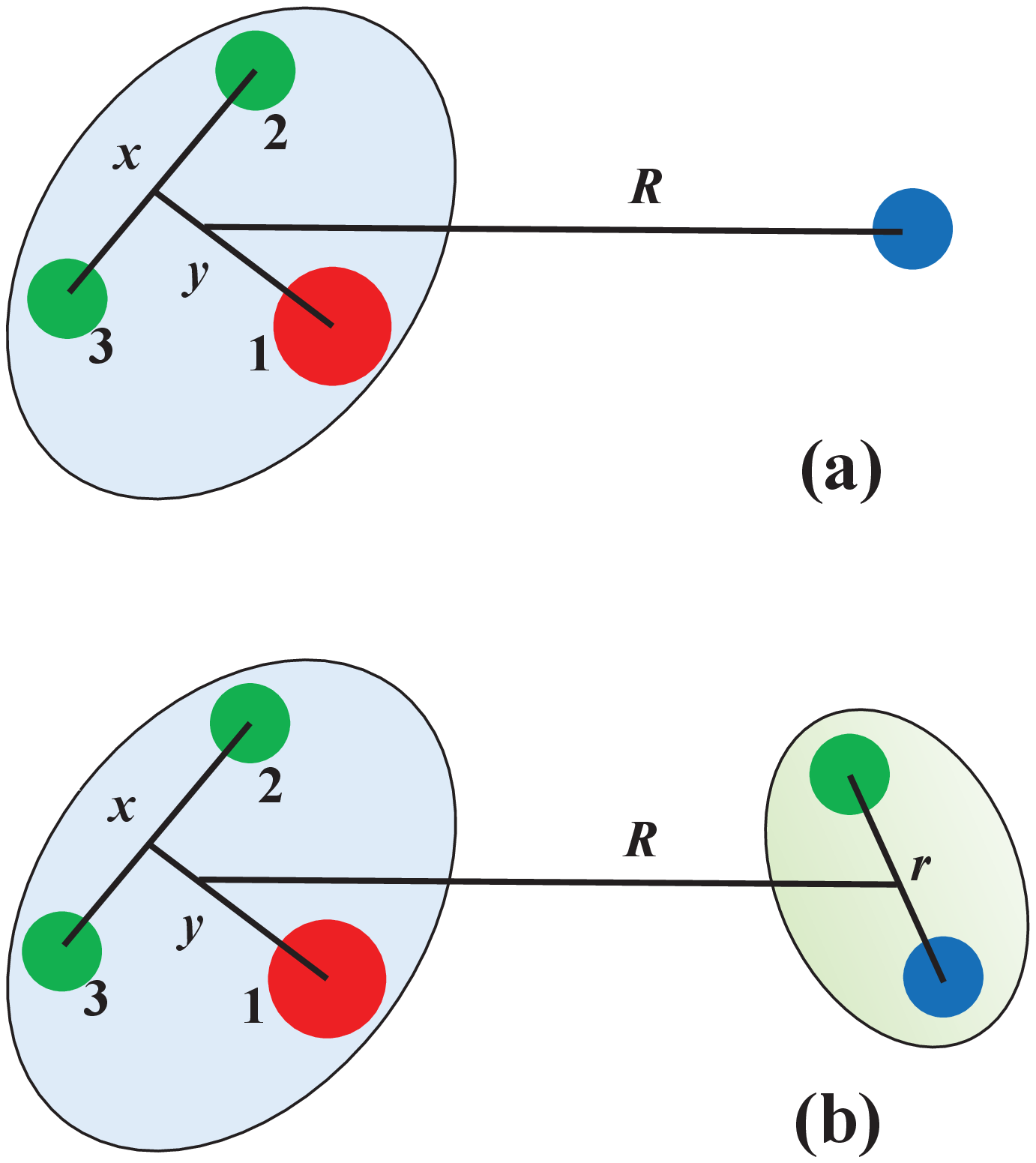}
\caption{Coordinates for the $\carp$ system (a) and for the $\card$ system (b). }
\label{jacobi}
\end{figure}

Figure \ref{jacobi} gives a schematic representation of the $\carp$ and $\card$ systems.
We define $\bm {\xi}_i$ as the internal coordinates of the two interacting nuclei ($ \pmb{\xi}_i = {\pmb r}_i$, $({\pmb x,\pmb y})$ for the two and three-body systems, respectively). Their internal Hamiltonian are
denoted as $H_i$, and the relative coordinate as ${\pmb R}$.

Considering $T_R$ as the relative kinetic energy, the Hamiltonian for the projectile + target system is written as
\begin{align}
H = H_1(\pmb{\xi}_1) + H_2(\pmb{\xi}_2) + T_R +\sum_{ij} V_{ij}(\bm \xi_1,\bm \xi_2,\pmb R), \label{e11}
\end{align}
where $V_{ij}$ represent two-body optical potentials between the fragments. In the present work, $V_{ij}$ contains $^{14}$C + $p$ and $n+p$ potentials for the $\carp$ scattering whereas for the $\card$ case it also contains $^{14}$C + $n$ and $n+n$ interactions. 

In the CDCC method, the total wave function of the system is expanded as
\begin{align}
\Psi^{JM\pi} = \sum_{\beta L I} u_{\beta L I}^{J\pi}(R)\,\varphi_{\beta L I}^{JM\pi}(\bm \xi_1,\bm \xi_2,\Omega_R), \label{e14}
\end{align}
where $L$ and $I$ represent the relative angular momentum and the channel spin, respectively. The index $\beta$ is defined as $\beta = (j_1,k_1,j_2,k_2)$, where $j_i$ and $k_i$ are the spins and excitation levels of nucleus $i$
(the parity is understood). In practice, the summation over $j_1,j_2$ and $k_1,k_2$ are truncated at some limiting values $\jmax$ and $k_{\rm max}$, which could be different for the projectile and for the target. The channel functions $\varphi_{\beta L I}^{JM\pi}$ are defined as
\begin{align}
\varphi_{\beta L I}^{JM\pi}(\bm \xi_1,\bm \xi_2,\Omega_R) = \Big[\big[\Phi_{k_1}^{j_1}(\bm \xi_i) \otimes \Phi_{k_2}^{j_2}(\bm \xi_2) \big]^I \otimes Y_L(\Omega_R)  \Big]^{JM}, \nonumber\\
\label{e15}
\end{align}
where $\Phi_{k_i}^{j_i}$ is the wave function of the colliding nucleus $i$ and can be obtained by solving the Schr{\"o}dinger equation 
\begin{align}
H_i \Phi_{k_i}^{j_i m_i \pi_i} = E_{k_i}^{j_i \pi_i} \Phi_{k_i}^{j_i m_i \pi_i}. 
\label{e16}
\end{align}
$E_k^{j\pi} < 0$ correspond to physical states, whereas $E_k^{j\pi} > 0$ correspond to PS. For the
proton, the internal wave function is of course unity, and the internal energy is zero.

The radial wave functions $u_{\beta L I}^{J\pi}(R)$ in Eq.\ (\ref{e14}) are solutions of the coupled differential equations
\begin{align}
&\Big(-\frac{\hbar^2}{2\mu}\Big[\frac{d^2}{dR^2}-\frac{L(L+1)}{R^2}\Big]+E_{k_1}^{j_1}+E_{k_2}^{j_2}-E\Big)u_{\beta L I}^{J\pi}(R) \nonumber\\
&+ \sum_{\beta' L' I'} V_{\beta L I,\beta' L' I'}^{J\pi}(R) u_{\beta' L' I'}^{J\pi}(R) = 0, 
\label{e18}
\end{align}
where the coupling potentials $ V_{\beta L I,\beta' L' I'}^{J\pi}(R)$ are given by
\begin{align}
 V_{\beta L I,\beta' L' I'}^{J\pi}(R) = \langle \varphi_{\beta L I}^{JM\pi}|\sum_{ij}V_{ij}|\varphi_{\beta' L' I'}^{JM\pi} \rangle, \label{e19}
\end{align}
which involves integrations over $\bm \xi_1$, $\bm \xi_2$ and $\Omega_R$. The calculations of these coupling potentials are given in the appendix of Ref.\ \cite{De20} for the 3 + 1 and 3 + 2 body systems. 

In practice, Eq.\ (\ref{e18}) may involve several thousands of coupled equations for each $J\pi$ and this represents the most challenging part of the CDCC calculations. However, the use of $R$-matrix along with the Lagrange-mesh method \cite{DB10,De16a} provides fast numerical computations and makes them feasible. With this approach we calculate the scattering matrices, which then provide the elastic, inelastic and breakup cross sections.

\section{$\carp$ scattering}
\label{sec4}
\subsection{Conditions of the calculations}
We calculate the $\carp$ and $\card$ elastic and inelastic scattering cross sections at a $^{16}$C energy of 24 MeV/nucleon, which corresponds to $\ecm = 22.59$ MeV for $\carp$ and to $\ecm = 42.67$ MeV for $\card$. Experimental data for these reactions have been recently published in Ref.\ \cite{JLY20}. 
We first discuss the case of $\carp$ which is simpler than the $\card$ scattering since breakup effects are present in $\car$ only.

Before presenting the cross sections, it is important to mention the conditions of calculations, which include $R$-matrix and Lagrange-mesh \cite{DB10,De16a} parameters, various potentials and parameter $K_{\rm max}$ for various $J$ values.
For the $R$-matrix method, we use a channel radius $a = 25$ fm and 50 Lagrange basis functions which guarantee a good convergence of the calculations. Small changes in these parameters do not bring any significant modification in the cross sections. Large channel radii need more basis functions which increases the computation times. Optimizing the choice of the channel radius is therefore an important issue.    

For $\carp$, we need two optical potentials: we use the Minnesota interaction \cite{TLT77} for $n+p$ and the Koning-Delaroche (KD) global potential \cite{KD03} for $^{14}{\rm C}+ p$. Additionally, we also perform the calculations using the Chapel Hill (CH) parametrization \cite{VTM91} for the $^{14}{\rm C}+ p$ interaction, which allows us to assess the sensitivity of the cross sections to this optical potential.
 
We have considered $j = 0^+, 1^-, 2^+$ and $3^-$ PS of $^{16}$C up to a maximum energy $\emax=20$ MeV, which are calculated using the procedure described in Sect.\ \ref{sec2}. In fact, a good convergence is already achieved with $\emax = 10$ MeV. For these calculations, we use $K_{\rm max}=16$, which provides converged $^{16}$C energies and keeps the number of PS within reasonable limits. A maximum angular momentum of $J_{\rm max}=25$ is used to compute the cross sections. We have performed various tests against all these parameters to ensure the convergence of the calculations. In particular, we ensure that the cross sections does not vary by more than $1-2 \%$ while changing these parameters beyond a certain value.

\subsection{$\carp$ elastic cross section}
\begin{figure}[ht]
\centering
\includegraphics[width=8.6cm]{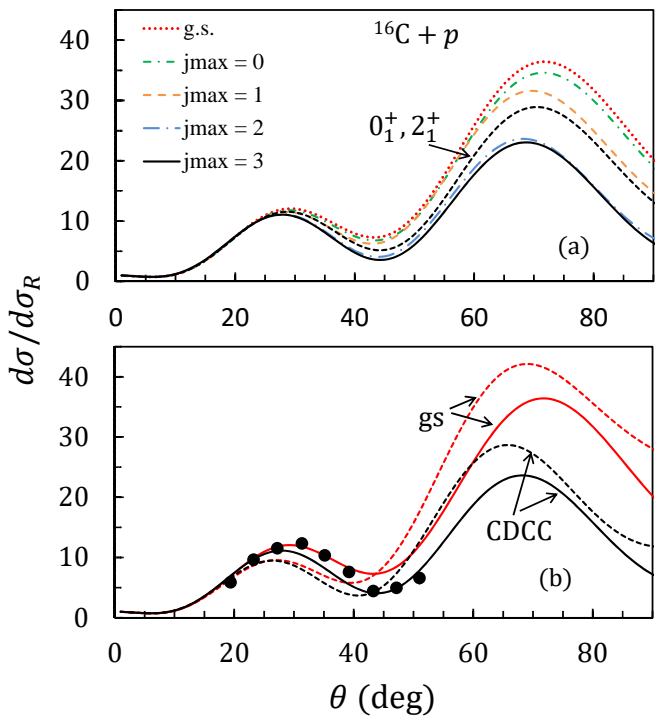}
\caption{$\carp$ elastic scattering cross sections divided by the Rutherford cross sections, at $\elab =24$ MeV/nucleon energy of $^{16}$C ($\ecm=22.59$ MeV). (a) Convergence with respect to $j_{\rm max}$. (b) Comparison with the experimental data of Ref.\ \cite{JLY20} using the KD (solid lines) and CH (dashed lines) $^{14}$C + $p$ potentials. }
\label{16Cp}
\end{figure}

In Fig.\ \ref{16Cp}, we plot the ratio of the elastic scattering to the Rutherford cross sections for $\carp$ at $\ecm= 22.59$ MeV and compare them with the experimental data of Ref.\ \cite{JLY20}. In Fig.\ \ref{16Cp}(a), we check the convergence of the cross sections with respect to  $j_{\rm max}$. It is clear from the figure that the contribution of $j=1^-$ pseudostates is small, whereas $j=2^+$ PS are the most important. This is explained by the presence of the $2^+_1$ first excited state. Calculations with only $0_1^+$ and $2_1^+$ states of $^{16}$C are not very different than the full calculations (with $j_{max}=2$)  from $0^\circ$ to $60^\circ$, although some difference can be seen at larger angles which indicates the importance of non-resonant continuum at larger angles. Also it can be seen that cross sections for $j_{\rm max}=3$ are not much different than for $j_{\rm max}=2$ which confirms the convergence of the calculations. Another important information one can collect from this figure is that at this energy, breakup effects are insignificant for $\theta<25^\circ$. 

In Fig.\ \ref{16Cp}(b), we compare the CDCC cross sections with the data. The calculations involving the $^{16}$C ground state only overestimate the experimental cross section for $\theta>40^\circ$ with both potentials. On the other hand, the solid line which corresponds to the full CDCC calculation with the KD potential nicely agrees with the data, except in the range $\theta \sim 30^\circ-40^\circ$, where the model slightly underestimates the experimental data. This shows that breakup effects are important for $\theta>40^\circ$. The cross section computed with the CH parametrization for $^{14}$C + $p$ which are less good than with
the KD potential. This shows that a proper knowledge of $^{14}$C + $p$ potential is important for these calculations. 

\begin{figure}[ht]
	\centering
	\includegraphics[width=8.6cm,trim=0.0cm 0.0cm 0.0cm -1.0cm,clip]{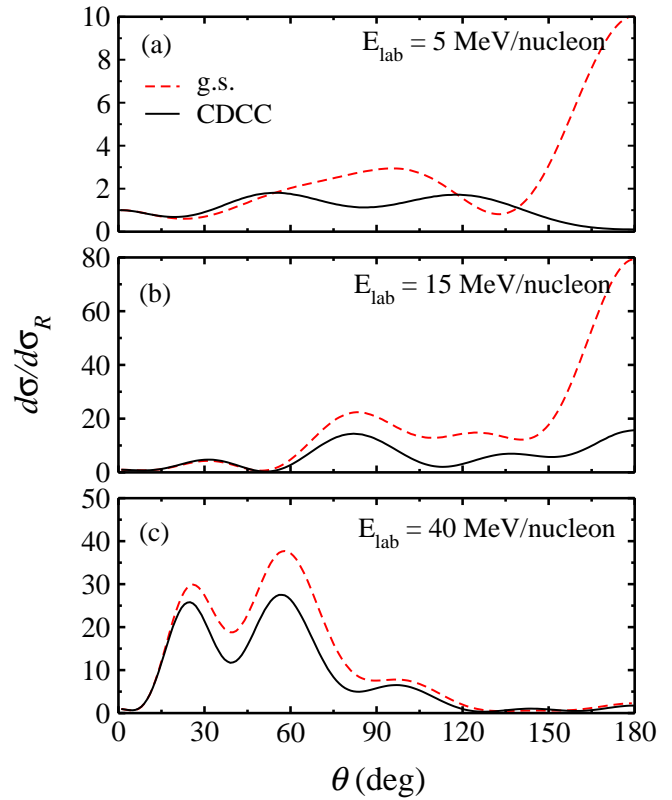}
	\caption{$\carp$ elastic scattering cross sections divided by the Rutherford cross sections at four  $^{16}$C energies. Dashed and solid lines represent calculations performed with only the g.s. of $^{16}$C and with the full CDCC model, respectively. }
	\label{16Cp_eng}
\end{figure}

We also apply the CDCC model to predict cross sections at other energies. In Fig.\ \ref{16Cp_eng}, we plot  the elastic cross sections at three different beam energies of $^{16}$C which are 5, 15 and 40 MeV/nucleon ($\ecm=$ 4.71, 14.12, 37.65 MeV, respectively) using KD potentials. Keeping all the other conditions and parameters unchanged, we have performed the CDCC calculations and compare them with the single channel case. We conclude that going from low to higher energies, the difference between the two calculations shift from higher to lower angles. Furthermore, this difference itself decreases as one moves to higher energies. 

\section{$\card$ scattering}
\label{sec5}
\subsection{Conditions of the calculations}
Now we discuss the $\card$ scattering for which the calculations are more complex and time consuming than in the $\carp$ scattering. This is due to the larger number of channels involved in $\card$. We take the breakup channels of deuteron also into account due to its low breakup threshold. Furthermore, as discussed in Ref.\ \cite{De20}, the coupling potentials (\ref{e19}) for the 3 + 2 body systems are more complex and involve multi-dimensional integrals. Therefore it is quite difficult to achieve the full convergence of the cross sections over a wide angular range. 

For the feasibility of the full calculations we take $j_{\rm max} = 2$ for $^{16}$C and deuteron partial waves are considered up to $j_{\rm max} = 4$. In these calculations, most of the conditions are the same as for the proton target but to decrease the number of channels, $K_{\rm max} = 12$ and $N=15$ Gauss-Laguerre basis functions are used [in Eq.\ (\ref{e10})]. This decrease does not bring any noticeable change in the cross sections. Furthermore, PS up to $E_{\rm max} = 8$ MeV are considered for the $^{16}$C as these are enough to achieve the satisfactory convergence whereas for the deuteron we considered PS up to $E_{\rm max} = 20$. In fact, increasing  $E_{\rm max}$ from 20 to 30 MeV for the deuteron slightly decreases the cross sections in the angular range from $60^\circ$ to $120^\circ$ and almost no change at other angles, which again ensures the convergence of the calculations.  To calculate the PS in deuteron we use 20 Lagrange basis functions (Gauss-Laguerre) with a scaling parameter $h=0.3$ fm (see for example Ref.\ \cite{Baye15} for more detail). 

For the $\carn$ interaction, we use the KD potentials and as we did in the previous case. Here also we test the sensitivity of the calculations by using the CH interaction. For the $n+n$ and $n+p$, we use the Minnesota potential \cite{TLT77}.

\subsection{$\card$ elastic cross section}
\begin{figure}[ht]
	\centering
	\includegraphics[width=8.6cm,trim=0.0cm 0.0cm 0.0cm 0.0cm,clip]{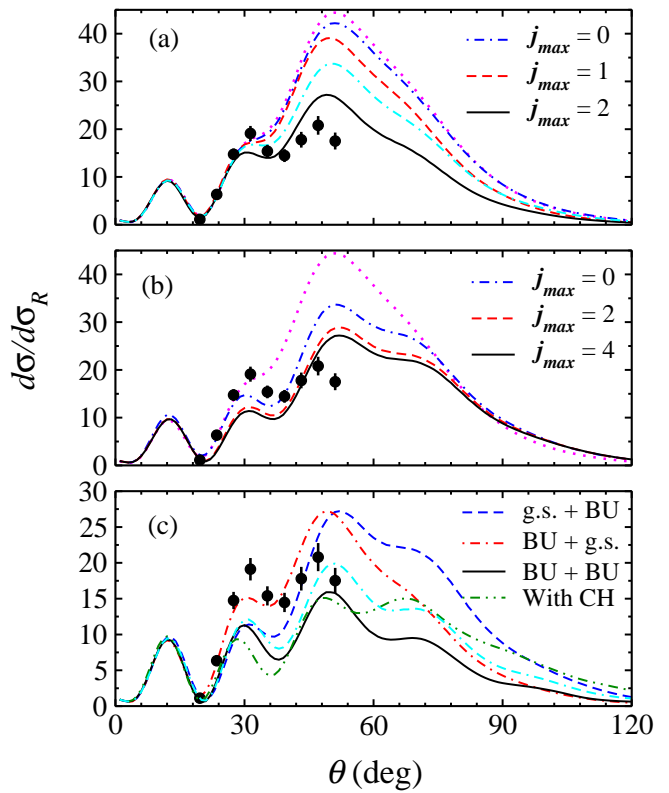}
	\caption{$\card$ elastic scattering cross sections divided by the Rutherford cross sections at $E_{lab} =24$ MeV/nucleon energy of $^{16}$C ($E_{c.m.}=42.67$ MeV). (a) only $^{16}$C breakup is included (b) only $d$ breakup is included (c) convergence of full five-body calculations. The dotted lines in panels (a) and (b) represent single-channel calculations where only the ground states of $d$ and $^{16}$C are included, whereas, double-dashed-dotted lines in panels (a) and (c) are calculations performed with only the $0_1^+$ and $2_1^+$ states of $^{16}$C. Experimental data are taken from Ref.\ \cite{JLY20}. }
	\label{16Cd}
\end{figure}

In Fig.\ \ref{16Cd}, we plot the $\card$ elastic cross sections. We first consider the breakup in one particle at a time before performing the full $3+2$ body CDCC calculations. In Fig.\ \ref{16Cd}(a), we consider the breakup of $^{16}$C, whereas the deuteron is in the ground state. For a comparison we also plot the single-channel cross sections (dotted line), where only the g.s. of $^{16}$C and of $d$ are included. It is evident that single channel calculations are unable to explain the data for $\theta>30^\circ$. Including the continuum in $^{16}$C reduces the magnitude of the cross sections, as in $\carp$ case. One can see that $j=2^+$ PS significantly change the cross sections whereas those with $j=1^-$ have a small influence. 

In Fig.\ \ref{16Cd}(b), we consider breakup channels in the deuteron, whereas $^{16}$C is in its ground state. The convergence with respect to $j_{\rm max}$ is clear. Again, increasing $j_{\rm max}$ decreases the magnitude of the cross section. However, neglecting $\car$ breakup leads to small differences in the peaks near $\theta \approx 30^{\circ}$ and $\theta \approx 50^{\circ}$. 

In Fig.\ \ref{16Cd}(c) we plot the full five-body calculations when breakup effects are included in $^{16}$C as well as in $d$ (solid line). For comparison, we also plot the other two possibilities considered in Figs. \ref{16Cd}(a) and (b). As mentioned earlier, full calculations are quite challenging. We deal with a total of 504 channels. It is clear from the figure that although the shape of the data is reasonably well reproduced, the magnitude of the cross sections is underestimated in the angular range $30^\circ-60^\circ$.
In Fig. \ref{16Cd}(c) we also compare the five-body calculations performed by using the CH optical potentials for $\carp$ and $\carn$. One can see a difference especially at larger angles ($>60^\circ$), but this difference is smaller than breakup effects. 

We also investigate the importance of $2_1^+$ state of $^{16}$C in these calculations. Double-dashed-dotted lines in Fig.\ \ref{16Cd}(a) and Fig.\ \ref{16Cd}(c) are calculations performed with only the $0_1^+$ and $2_1^+$ states of $^{16}$C. It is clear that these calculations are not very different than the full calculations (solid lines) in both these figures, especially in the angular range of the available data (as in $\carp$ system) although at larger angles non-resonant continuum plays some role.  We further found that other bound states ($0_2^+$ and $2_2^+$) of $^{16}$C have negligible influence on the cross sections. This can be seen in the context of deeply bound nature of  $^{16}$C.

\begin{figure}[ht]
\centering
\includegraphics[width=8.6cm,trim=0.0cm 0.0cm 0.0cm 0.0cm,clip]{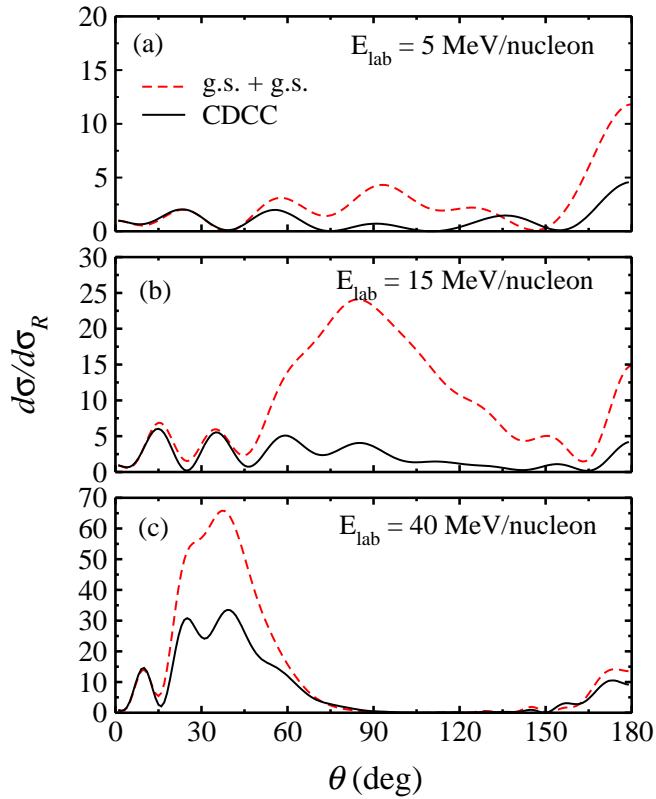}
\caption{$\card$ elastic scattering cross sections as a ratio to the Rutherford cross sections at different energies of $^{16}$C. Dashed lines in each panel represent single channel calculations performed with only the g.s. of $^{16}$C and $d$ whereas solid lines represent five-body CDCC calculations.  }
\label{16Cd_eng}
\end{figure}

As for the $\carp$ system, we also perform the calculations to predict cross sections at some other energies of $^{16}$C which again we consider as 5, 15 and 40 MeV/nucleon and they correspond to $E_{c.m.}$ of 8.89,  26.67, 71.11 MeV, respectively. We have kept all the conditions unchanged. In Fig.\ \ref{16Cd_eng}, we plot these cross sections (solid lines) and compare them with single channel case (dashed lines).  Again, we can see that with increase in energy, the amplitude of the difference between CDCC and single channel calculations shift to the lower angles. Furthermore, it shows that the breakup effects are relatively more stronger at medium energies. This can be expected as at higher energies the interaction time will be relatively small than at medium energies, whereas at low energies particles may not come close enough to interact strongly.

\section{Inelastic cross sections}
\label{sec6}
Various methods have been used in the literature to determine the $E2$ transition probability, and there
is still a large uncertainty. The inelastic cross sections to the first $2^+$ state of $\car$ has been measured in Ref.\ \cite{JLY20}, and
used to determine the $E2$ transition probability from an optical-model analysis involving a deformation parameter
$\delta$. The fitted value $\delta=1.07 \pm 0.26$ fm was then converted to 
$B(E2)=4.34^{+2.27}_{-1.85}\ e^2.{\rm fm}^4$.

Core $+n+n$ three-body models, however, are known to underestimate this
transition probability since core-deformation effects are in general absent. As shown in Sec.\ \ref{sec2}.C, this value can be reproduced by the $\carnn$ model provided that an effective charge $\delta e \approx 0.3 e$ is used. Notice that,
owing to the small charge of the target ($Z=1$), the non-monopole Coulomb interaction (proportional
to the $E2$ transition amplitude) plays a minor role, and has been neglected in the analysis \cite{JLY20}.

\begin{figure}[ht]
\centering
\includegraphics[width=8.6cm]{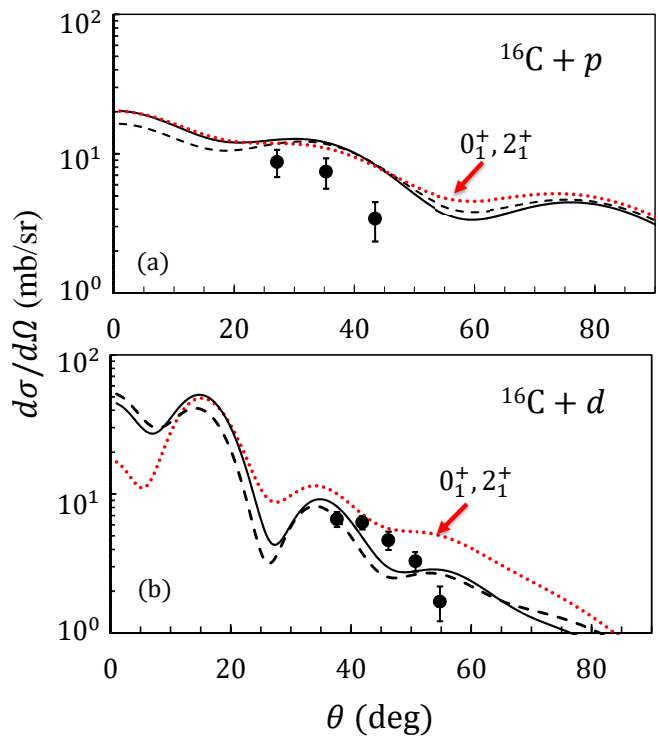}
\caption{Angular distributions of inelastic scattering to the $2_1^+$ state of $^{16}$C on a (a) proton and (b) deuteron target at $E_{lab} =24$ MeV/nucleon. Experimental data are taken from Ref.\ \cite{JLY20}. Dotted lines in both panels represent calculations when we take only the $0^+_1$ and $2^+_1$ states of $^{16}$C, whereas solid  and dashed lines correspond to the full CDCC calculations performed with the KD and CH optical potentials for $^{14}$C + nucleon.}
\label{inela}
\end{figure}

In Fig.\ \ref{inela}, we plot the angular distributions of inelastic scattering on proton (a)
and  deuteron (b) targets, and compare them with the data from Ref.\ \cite{JLY20}. Calculations are performed in the CDCC framework considering $3+1$ and $3+2$ body configurations, respectively with the KD (solid lines)
and CH (dashed lines) potentials. It can be seen that calculations performed with these two different potentials, give nearly the same results in both cases over the considered angular range. 

For a comparison, we also perform calculations using just the $0^+_1$ and $2^+_1$ states of $^{16}$C and for the deuteron target we also consider the ground state only. These calculations show that, for the proton target, breakup effects in $^{16}$C does not have much influence on the inelastic cross sections, although they slightly improve the shape of the angular distribution in the angular range $40^\circ-50^\circ$.
On the other hand, for the deuteron target, the inclusion of breakup effects improves the calculations. They are important to explain the data especially for $\theta > 50^\circ$.

\begin{figure}[ht]
\centering
\includegraphics[width=8.6cm, trim=0.0cm 0.0cm 0.0cm 0.0cm,clip]{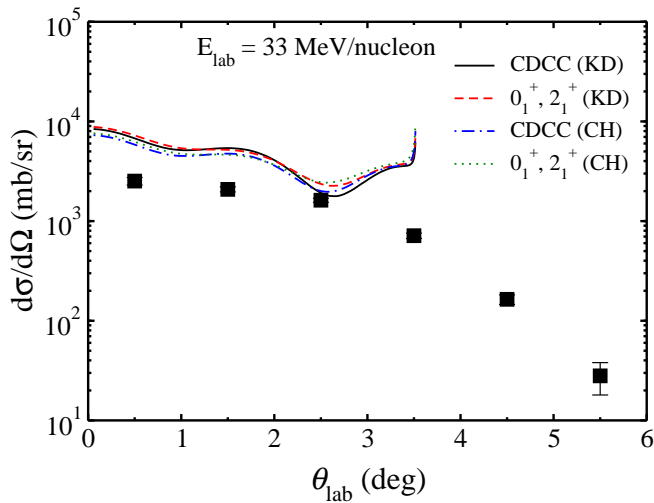}
\caption{Angular distributions of inelastic scattering to the $2_1^+$ state of $^{16}$C on a proton target at $E_{lab} =33$ MeV/nucleon. Experimental data are taken from Ref.\ \cite{OIA06}. Solid and dot-dashed lines are full CDCC calculations performed with the KD and CH optical potentials, whereas dashed and dotted lines represent corresponding calculations performed with only the $0^+_1$ and $2^+_1$ states of $^{16}$C. Calculations are converted to the lab frame in order to be consistent with the data but due to the reaction kinematics $\theta_{lab}$ is confined between  $0^\circ-3.6^\circ$.}
\label{inela33}
\end{figure}

Additionally, we also perform $3+1$ body calculations to calculate the inelastic cross sections to the first $2^+$ state of $\car$  at 33 MeV/nucleon. We plot these cross sections in the lab frame in Fig. \ref{inela33}, using both the KD (solid line) and CH (dot-dashed line) potentials and compare them with the data of Ref.\ \cite{OIA06}. Dashed and dotted lines are corresponding calculations when we consider only the ground and $2_1^+$ states of $^{16}$C. Reaction kinematics limits our calculations within $\theta_{lab}=3.6^\circ$ and full CDCC calculations are not very different than those with only the $0^+_1$ and $2^+_1$ states of $^{16}$C, which is consistent with Ref.\ \cite{KO19}. However,  our calculated cross sections up to around $2^\circ$ are nearly double to those reported in Ref.\ \cite{KO19} where $p-^{16}$C potential was microscopically derived by folding the Melbourne $g$-matrix $NN$ interaction with the target densities obtained from the antisymmetrized molecular dynamics. This again indicates a need for proper potential at this energy.

\section{Conclusion}
\label{sec7}
The goal of the present work is the study of $\carp$ and $\card$ scattering, by including breakup effects. The
$\car$ nucleus is described by a $\carnn$ three-body configuration and its breakup is simulated by pseudostates.
In $\card$, the target $d$ is defined by a $p+n$ structure, and pseudostates are also included.
This leads to very demanding calculations, since the total number of channels are the product of $\car$
and of $d$ states. This can be achieved, however, with modern computer facilities.

In $\carp$, we have shown that a fair agreement with the recent data of Ref.\ \cite{JLY20} can be 
obtained. Breakup effects are not strong, but improve the agreement with experiment for $\theta>40^{\circ}$. As a general statement, the availability of data at large angles would be extremely
helpful to assess the models. We have shown that the sensitivity to breakup effects increases at large angles.

The $\card$ elastic scattering is reasonably well reproduced by the five-body CDCC model, considering that
there is no adjustable parameter. Our results suggest that both the $\car$ and deuteron breakups have
an influence on the elastic scattering cross section. This confirms a previous conclusion on
$^{11}{\rm Li}+d$ scattering \cite{De20}. However, due to the deeply bound nature of $\car$ as compared to $^{11}{\rm Li}$, non-elastic effects below $60^\circ$ are mainly contributed by the $2_1^+$ state of $\car$. 

Although the $B(E2)$ value in $\car$ is small in the three-body model without effective charge,
the inelastic cross sections are in reasonable agreement with experiment. This stems from the low influence of
the Coulomb interaction for light targets. The inelastic cross sections are therefore mainly sensitive to nuclear
effects.

\section*{Acknowledgment}
This work has received funding from the European Union's Horizon 2020 research and innovation program under the Marie Sk{\l}odowska-Curie grant agreement No 801505. It was also supported by the Fonds de la Recherche Scientifique - FNRS under Grant Numbers 4.45.10.08 and J.0049.19. It benefited from computational resources made available on the Tier-1 supercomputer of the 
F\'ed\'eration Wallonie-Bruxelles, infrastructure funded by the Walloon Region under the grant agreement No. 1117545. P.D. is Directeur de Recherches FNRS.


%

\end{document}